# Precarious Experiences: Citizens' Frustrations, Anxieties and Burdens of an Online Welfare Benefit System

Colin Watson

Open Lab, Newcastle University, c.watson8@newcastle.ac.uk

Adam W Parnaby

Open Lab, Newcastle University, a.w.parnaby2@newcastle.ac.uk

Ahmed Kharrufa

Open Lab, Newcastle University, ahmed.kharrufa@newcastle.ac.uk

There is a significant overlap between people who are supported by income-related social welfare benefits, often in precarious situations, and those who experience greater digital exclusion. We report on a study of claimants using the UK's Universal Credit online welfare benefit system designed as, and still, "digital by default". Through data collection involving remote interviews (n=11) and online surveys (n=66), we expose claimants' own lived experiences interacting with this system. The claimants explain how digital channels can contribute to an imbalance of power and agency, at a time when their own circumstances mean they have reduced abilities, resources and capacities, and where design choices can adversely affect people's utility to leverage help from their own wider socio-technical ecosystems. We contribute eight recommendations from these accounts to inform the future design and development of digital welfare benefit systems for this population, to reduce digital barriers and harms.

## 1 INTRODUCTION

The welfare state in many jurisdictions provides financial help to those on low incomes, often requiring systems to check eligibility and level of award. Previous work [92] shows working-age citizens can be distressed and overwhelmed making non-digital benefit applications, negatively affecting their mental and physical health and wellbeing, and identifies a need to reduce complexity, so claimants retain some degree of control. Increasingly such services are being digitised in numerous countries through e-government initiatives [1,18,21,26,57,81]. Accordingly, our study examines people's first-hand experiences of applying for and maintaining their benefit awards digitally. We do this using a social welfare benefit in the UK called Universal Credit (UC) [30], which is delivered as a periodic financial payment to help with living costs. This type of citizen support is a type of social protection provided as a cash payment [39,52,64]. UC was designed and developed over the last ten years [32,66], by the UK State's Department for Work and Pensions (DWP), and currently provides support to 5.6 million working-age citizens ("claimants"), predominantly accessed online [3,33]. UC targets those in poverty through income and capital eligibility requirements, providing a payment to those who are unemployed, those working on low incomes, and those unable to work, for example due to disability. The DWP has a monopoly on the provision of this public service and has two roles: the administration of welfare benefit claims and supporting claimants into work and more or better work. These are reflected in a range of personalised ongoing commitments (responsibilities) claimants have to agree to do (e.g.: to prepare for and look for work, to increase earnings if in work, to attend telephone appointments, to report changes in circumstances) to receive UC payments and to prevent being sanctioned (which entails benefit payment deductions).

First and second order digital divides [27,74] are often related to people's backgrounds and socio-economic status [45] and UC claimants have precarious [19] lives related to unemployment and underemployment, with a significant proportion having disabilities or long-term health conditions [31]. These are also groups associated with being less digitally skilled [1,25,72,83,94] and more at risk from digital design marginalisation [84], whose unjust digital experiences [28] are less recognised than other groups. Additional aspects that make this particular welfare system interesting from a HCI viewpoint are 1) that it was planned as "digital by default"; 2) that many of the interactions with claimants are driven by a conditionality regime, backed by various sanctions including financial penalties; and 3) the lack of transparency in its design and operation. Moreover, the system is impermeable with no external digital integration capabilities.

Our study concerns claimants' "technology-mediated" [21] interactions with digital systems, not the particular jurisdiction-specific policies enacted. Additionally, our study does not report on any form of automated decision-making [51], which is not an aspect of UC's current design apart from for fraud prevention [3,18]. Millar and Whiteford [61] have described the reality of the currently deployed system as being built on assumptions of how people live their lives, rather than the reality of people's lives. Claimants have been found to be burdened through a system "beyond their control", especially the more vulnerable [73]. Many of UC's policy choices are being debated and challenged by others (e.g.: [2,20,34,60,93]), but there has been less attention on the digital implementation aspects. Our study examines what it is about UC's remote asynchronous channel (UC Online) that contributes to *digital barriers and harms* to citizens, and how these might be mitigated.

With 68 of our 77 unique participants still currently claiming UC, for an average of 22 months, our study centres the views of claimants through two rounds of interviews, and a survey. Our contributions are twofold: 1) insights from the perspective of a group of people using UC Online; and 2) design implications for those designing digital systems and other interventions, regardless of the technology ownership, relating to supporting claimants' wider socio-technical ecosystems, acknowledging claimants as people, and reducing the burdens of system interactions on claimants.

## 2 BACKGROUND AND RELATED WORK

### 2.1 Public administration and delivery of digital services

Delivery of State policies imposes administrative burdens on citizens [11] including costs of learning, compliance and psychological costs [16,47], which increase when eligibility and commitments need to be verified [86]. E-government aims to deliver policy more effectively, but the associated advantages and risks are not evenly spread through all the stakeholders and user groups [59,68]. Alston [4] describes how cost reduction, fraud reduction, work incentivisation and increased responsibilities, are often combined with erosion of human rights, and can disproportionately disadvantage already marginalised groups such as people with disabilities and health conditions, those with children, those who do not speak the State's primary language(s), and women [3,18,26,29,78]. Opportunities for improvision and discretion are reduced [70]. Potential advantages to citizens (e.g. greater convenience [43], better access to knowledge and information [67], more discrete contact [14,43], reduced judgemental discretion bias [1]) are less clear for poor people [19], and have led to reduced service adoption and greater need for support from people's own social networks [48]. Harris [46] identifies how digital provision places more significant burdens on those in poverty, with Coles-Kemp and Jensen [19] highlighting how people perceive a need to be online and available at almost any time, and Sin et al. [84] showing that experiencing digital marginalisation is a burden in itself. Additional burdens require greater use of cognitive capacity, which is more limited when people are in unstable circumstances such as unstable work, debt, poor mental health and monetary scarcity [5,42,80,82]; these are the same precarious situations many welfare benefit policies seek to mitigate. Eubanks [38] explains how the poor can be adversely targeted in digital systems, by profiling, surveillance, and punishment, reducing their self-determination and agency. Graeber describes how public administration excels at making generalisations, failing to understand people's own complex and ambiguous existences [41], which Birhane [6] likewise describes as "ambiguous, indeterminable, and inherently unpredictable". Morris et al. [62] in common note how the simplifications and efficiencies of e-government have to cover up the "messiness" of centralised welfare services provision where Considine et al. [21] suggest "efficiency and consistency" need to be balanced with "personalisation, flexibility and inclusion". Additionally, these new technologies may have to coexist forever with other channels [43], including street-level bureaucracy [53] and rely on non-digital intermediaries, who are already a significant part of the social infrastructure [17], for successful delivery [29,69,77]. Consequently, citizens also have to navigate complicated relationships between civil servants, digital infrastructure, other parties and themselves [10], creating potential for greater digital exclusion [85]. Therefore, given the close relationship between digital exclusion and social exclusion [27,72,74,84,94] where existing digital access [24] and skill [44] inequalities can be exacerbated by new forms of inequality [1,16], Massimi [58] suggests the promotion of productivity and efficiency should not be the role of technology for navigating severe life disruptions. He recommends poverty-targeted public services should instead seek to minimise burdens and thus cognitive effort [7,71].

## 2.2 Universal Credit

When it comes to UC, which we use in our study, the implementation has been found to move complexity onto claimants [61,87] which people experience as "dehumanizing, punishing and humiliating" affecting their physical and mental wellbeing [63,78]. De Oliveira [65] also describes a lack of accountability by the DWP. Alston [3] raised concerns UC is a barrier to accessing benefits. While he cites many policy-related causes, others such as internet access, digital skills and support, data delays and potential automated risk detection for fraud and error are highlighted as digital channel related concerns. Our research question narrows our interest to how the primarily digital-only implementation of UC itself contributes to citizens' experiences. In recent UC-related work, Coles-Kemp et al. [18] imagine reframing digital welfare benefit systems within communities to better balance values of "efficiency, employment and self-efficacy" with "solidarity, reciprocity and care". Alessiato [1] highlights the need for trust and cooperation, with Scullion and Curchin [78] identifying trust being dependent on better consistency and transparency in UC. It has been reported how the implementation of UC has transferred claimant support from the DWP to local organisations [15,88]; Morris et al. [62] compared centralised UC welfare provision with that offered at a local level by food banks, which identified three barriers to utilising "faceless" UC: access difficulties, lack of support and suspicion of the DWP. This was in contrast to food bank services built around and for their "service users" in an attempt to provide them with self-efficacy and dignity. Our study examines digital-specific causes and effects more deeply than Alston's concerns, building upon previous work by extending the understanding of claimants' online experiences, the challenges of using and getting help with such digital services, and how to address digital channel shortcomings.

### 2.3 HCI costs and benefits

In terms of HCI, Pei and Crooks [68] examined the overheads of people using digital technology at a community literacy centre in the United States. They describe the "startup, maintenance and affective and other costs" that arise with digital access, describing how the various burdens of access can be imposed both on the individual as well as their networks. Among the interaction design strategies for social justice defined by Dombrowski et al. [28] is "designing for distribution" where the benefits and burdens are more equitably available across the whole socio-technical system. In a study using Denmark's job placement system for unemployed people as the context, Holten Møller et al. [50] used a fictional app to investigate citizens' experiences of digital public services requiring them being more active data providers. They found a similar imbalance of costs and benefits when there is a "coercive exchange" of data needed to access the public service, and highlight how this is perceived as a lack of agency. Our own study explores how digital design choices perceived as beneficial to the State may transfer burdens to individuals, leading to inequitable distributions of benefits and burdens in operational systems (like in design as Dombrowski et al. [28]), and may also adversely affect people's networks and intermediaries who provide support (as Pei and Crooks [68]).

## 3 STUDY AND METHODOLOGY

Our research question for this work is: *what opportunities are there to change digital aspects of welfare benefit systems to counter the difficulties, barriers and related consequences experienced by claimants?* To answer this question, we undertook a study of UC Online exploring claimants' own experiences and views, primarily using remote semi-structured interviews and scenario-based research, backed up by an online survey. Undertaken over a 12-month period during the COVID-19 pandemic, commencing just after the most severe lockdown restrictions were eased in late summer 2020, we were not able to carry out other planned engagements such as in-person focus groups and workshops.

Claimants are considered a higher-risk population, and therefore we initially worked with professional staff from three local advice organisations (Stage A) to clarify the ethical issues, explore constraints and identify typical problems. Ethical approval was granted through the sponsoring university's review process.

### 3.1 Study design

Our study was completed in two stages as detailed below, with Stage A informing the claimant-focused Stage B.

<u>Stage A</u>: Foundational hour-long semi-structured remote interviews with professional participants (subsequently referred to as advisors) undertaken to gather their personal knowledge and experience, capturing information about their role, claimants' need for help, typical claimant technology access and benefit problems, difficulties providing assistance, and what changed during the pandemic. Advisor participants were not compensated for their time. Information from Stage A informed the questions asked in the first-round of interviews of Stage B.

<u>Stage B</u>: The study's core involving UC Online claimant participants, with two rounds of remote semi-structured interviews planned. The first-round gathered demographic and UC-related classification data, and asked questions about their own experiences of using the digital channel (and not benefit policy matters). The questions in this first-round were to inform the next round and related to: making the first claim (e.g.: "Did you get any help to make the claim? If so, what help, who from?"); using technology (e.g.: "Do you use a phone or something else to access the internet and UC"); using UC Online day-to-day (e.g.: "How often do you log onto UC?", "Have you ever been sanctioned? Why was that?"); getting assistance (e.g.: "Do you discuss UC with other people... queries, compare experiences and so on? Who?", "What would make claiming UC easier for you?"); changes during the pandemic; and an opportunity to mention anything else or ask questions. Participants received compensation for their time, at £10/hour, lasting mean 45min. The interviews were undertaken by telephone or online as participants preferred, with consent to record the audio.

In order to provide a more positive experience to participants [56] by sharing their ideas and demonstrating how their contributions were incorporated, even though they had no interactions with each other, scenario-based design [75] was used to convey four broad group ideas, drawn from the collective first-round data for use in the second-round of remote semi-structured interviews. As with [8,89], these scenarios were used as prompts (rather than answering specific questions [54]), each illustrating potential changes to UC Online, to offer wider options than they may have considered individually, encouraging further critical reflection thinking through topics raised by other participants. To counter presenting only positive viewpoints, and reduce participant response bias [23], the scenario documents also illustrated some less-obvious potential harms sourced using a HCI-focused hazard identification method [91] to systematically enumerate undesirable consequences. The resulting documents were reviewed by the third author, then by two independent external experienced researchers who have some familiarity with digital public services, and making adjustments after each review. Each scenario document comprised, on the first page, the matters being addressed (Figure 1a), then the visual collage-based scenario (Figure 1b), and on the last page three vignettes of potential harms identified (Figure 1c).

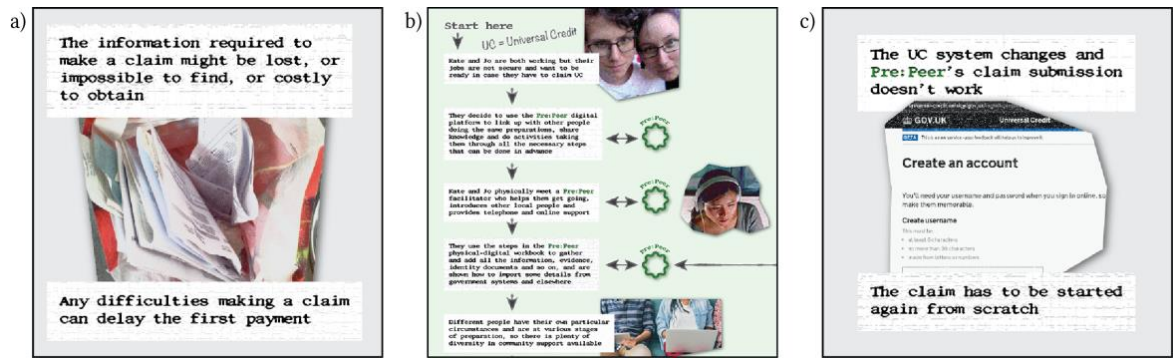

Figure 1: Extracts from one scenario used in the Stage B second-round interviews and surveys a) example claimant issue; b) part of the visual scenario showing changes; c) an example problem vignette.

These scenarios, selected to span the claim process from start to end and referred to later in the Findings, describe 1) community help to make an initial claim; 2) providing alternative communication channels; 3) delegated access; and 4) collaborative real-time shared viewing. These scenarios were discussed in the second-round interviews, which did not depend upon first-round interview participation. The identical set of questions was used for every scenario (e.g.: "What was your initial reaction to this idea when you first saw it?", "What is good about the proposal?'). These were also supplemented by questions about overall ranking of changes illustrated in the four scenarios (e.g.: "Which of these four proposals would be most useful to you and why?").

The second-round also involved collecting further demographic data (on computer anxiety, privacy concern and civic engagement) and information about UC status. Again, the interviews were undertaken by audio-recorded telephone/online, with a mean duration of 63min in addition to time taken reading the scenarios provided in advance, and compensated at the same rate as the first-round interviews.

The research team were unsure whether the data collection had reached saturation. Therefore a survey with a larger number of different UC Online claimants was undertaken to collect further data and assess whether this contributed to additional thematic analysis codes and/or themes. The survey was made available online and potentially on request in paper-format for those who prefer non-digital interactions. The survey used the scenario documents with identical questions to those in the second-round interviews with an additional open-ended question "Is there anything else you would like to add, good or bad about UC online, or your own technology ideas?". All participants received compensation for their time (at the survey platform's recommended rate of £7.50/hour); survey responses were collected by Google Forms and downloaded as textual data.

**3.2 Participants and recruitment**

Stage A: Advisors were recruited from local advice organisations which provided access to their staff, but otherwise did not shape the study [17]. The full-time advisors (n=3) had significant experience of UC and other UK welfare benefits. They assist citizens to use technology, and take up services related to welfare rights. Completed paper consent forms were scanned and emailed back to the research team.

Stage B: Due to the pandemic, the study did not involve any organisations who might have acted as gatekeepers providing access to their service users. Claimants were instead recruited through 1) letters posted to community organisations in both urban and rural settings; 2) approaching people who used the phrase

"universal credit" or "#universalcredit" on Twitter; and 3) through a poster campaign. Consent was recorded in an online form. For the second-round interviews, one participant dropped out and three further new participants became available to take part (first-round n=8, second-round n=10, total n=11). Different participants (n=66) were recruited for the survey by 1) posters; 2) letters to community groups; and 3) through the Prolific research study portal using a pre-screening survey to identify people who claimed UC online. Survey respondents all completed the consent stage online. A summary of all claimant participants across Stage B is provided in Table 1 indicating how survey participants tended to more female, younger and more likely to be working or looking for work, than interview participants.

Table 1: Demographic and UC claim characteristic comparison between all [a] interview and survey participants in study Stage B [b]

| Demographic/characteristic | | Remote Interview | Online Survey |
|---|---|---|---|
| Number of participants | n | 11 | 66 |
| Gender: Female / Male / other | ratio | 36% / 64% | 64% / 33% / 3% |
| Age grouped by decade *left* from 20s *to right* 60s | % | 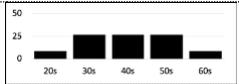 | 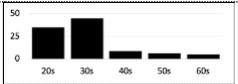 |
| Home location: Rural / Urban | ratio | 18% / 82% | 12% / 88% |
| Long-term sick or disabled | no (%) | 7 (64%) | 11 (17%) |
| Working part-time (1-29 hours/week) | no (%) | 1 (9%) | 23 (35%) |
| Working full-time (30+ hours/week) | no (%) | 1 (9%) | 5 (8%) |
| Had been on other benefits before UC | no (%) | 6 (55%) | 16 (24%) |
| Initial claim made during pandemic | no (%) | 5 (45%) | 26 (39%) |
| Needed help making initial claim for UC | no (%) | 3 (27%) | 12 (18%) |
| Still claiming UC at time of participation | no (%) | 11 (100%) | 57 (86%) |
| For those still claiming, duration of claim | months (mean) | 19 | 23 |

Notes a: First and/or second-round interviews and all survey; b: Stage A advisor participants are not included in comparisons.

### 3.3 Data analysis

The goal of data analysis was to understand issues with the UC Online welfare benefit system foregrounding the needs and concerns of claimants. Data from Stage A was only used to inform Stage B questions. As such, this data analysis is based on Stage B data only. The interviews were transcribed verbatim, following data cleaning to remove details identifying people or places. The resulting transcripts, and survey responses, were imported into NVivo qualitative data analysis software for management, coding, review and validation, using inductive thematic analysis [9]. Our analysis spanned data about the *existing system* and people's desire for changes, including such data prompted by the four scenarios (a separate analysis, not discussed in this paper, was undertaken to explore all the different types of harms and how the *design changes* in the scenarios add to or alter these, see [90]). The coding was undertaken by the first author in three steps: the first-round interviews; next the second-round interviews; and finally the surveys. Codes were added to, merged or modified during coding and earlier data re-coded as necessary. The survey data was also used to check (see Findings) whether data from the relatively small interview population sample had reached saturation or if the sample size needed to be increased [13]. The second author, with detailed knowledge of the digital exclusion subject area, undertook a complementary role through discussion of the codes, themes and their interpretation, leading to some recoding and additional interpretation insights.

## 4 FINDINGS

The thematic analysis of all interview transcripts and survey responses from Stage B conceptualised an overarching theme of *Life Servicing Capabilities* describing a view of public *service* delivery, where the combined *capabilities* of individual citizens, the wider community and the State need to come together in a timely manner for an individual's *life* needs. This overarching theme comprises the themes of *Potentiality*; *Temporality*; *Transparency*; and *Proficiency*, each of which include two sub-themes. Due to claimants' stated wariness of any adverse impacts on them by participating in our study, anonymous identifier labels are not presented adjacent to quotations below, since individuals could potentially be re-identified from each person's collective quotations if combined with State-held data. Table 2 provides a summary of theme coding versus participants: all sub-themes include codes assigned to every interview participant; there are no codes that only apply to survey response data indicating we had reached saturation, and no additional interviews were necessary. Although analysis of survey data did not generate new codes, it did contribute to further insights and quotations within existing themes.

Table 2: Thematic analysis coding comparison between the scenario-related interview transcripts and survey responses showing the number and percentage of participants coded to each sub-theme and the mean (M) number of extracts coded per participant

| Theme | Sub-theme | Remote Interview (n=10) | | Online Survey (n=66) | |
|---|---|---|---|---|---|
| | | n | Extracts | n | Extracts |
| Potentiality | Unregistered Users and Unauthenticated Users | 10 (100%) | M 13.1 | 63 (95%) | M 6.4 |
| | Active Authenticated Users | 10 (100%) | M 10.3 | 60 (91%) | M 4.9 |
| Temporality | Moment | 10 (100%) | M 12.2 | 63 (95%) | M 4.8 |
| | Rhythm | 10 (100%) | M 8.5 | 57 (86%) | M 4.1 |
| Transparency | Actions | 10 (100%) | M 10.3 | 56 (85%) | M 3.9 |
| | Ecosystem | 10 (100%) | M 22.0 | 61 (92%) | M 6.3 |
| Proficiency | Slips & errors | 10 (100%) | M 7.9 | 59 (89%) | M 3.4 |
| | Competence | 10 (100%) | M 7.3 | 41 (62%) | M 2.2 |

To help understand terminology used by claimants, UC Online's website has the following main parts: messaging area (Journal), list of actions (To Do List), list of commitments (duties and obligations), change of circumstance forms, and payment amounts. The vast majority of UC claimants are "online claimants" where most of their interactions are through UC Online, predominantly by the written message system of the Journal. There is a telephone helpline for these claimants, but this is not an alternative channel because it does not provide the functionality or features of UC Online, and is not staffed by decision-makers. Some face-to-face meetings can occur at high-street located "JobCentres" such as for initial identity verification processes, periodic meetings with work coaches (the latter also undertaken by telephone call). But in general the physical JobCentres are not part of the UC system for these online claimants. The term "agents" is used below to refer to all civil servants and others who work for, or on behalf of, the DWP.

### 4.1 Potentiality

This theme gathers references across fourteen codes to the existence, accessibility and usability of systems and processes to achieve citizens' goals. It encompasses digital access and capability matters, in two sub-themes: codes related to *Unregistered Users and Unauthenticated Users* of the service, and as *Active Authenticated Users*.

*4.1.1 Unregistered Users and Unauthenticated Users*

There is an awareness of <u>technology as a barrier</u>. A claim for UC can be unexpected, related to a disruptive life event, during what may be difficult personal circumstances e.g.: *"you're already in like a bit of a bad situation where you're already quite stressed already"* and *"I wasn't very well when I first originally went on to Universal Credit"*. Scenario 1 motivated one person to emphasize that without registering online *"you're locked out"*. Some people *"have limited access"* and *"had to get [broadband installed] so it was an extra expense"* and for others mobile data is expensive: *"cost of data"*, and may not be available reliably due to location or capacity. Barriers are greater for some e.g.: *"I know quite a few people who are not like IT literate... if they had to claim UC they would have big big problems"* such as *"people who genuinely need help such as learning disabilities"* or by those who would be helped by official *"translation capabilities"*, mentioned by five participants, due to a lack of trust in free translation tools.

<u>Inflexible identity verification</u> was an initial hurdle to register and use UC e.g.: "*I didn't have a passport... I didn't have a photo driving licence"* and *"I struggled proving my ID"*. Someone *"did have a [government] gateway ID... but at that point in time it wasn't acceptable"*. For most, the online verification failed, and required checking at a JobCentre, which is not always straightforward: *"I was in the JobCentre for nearly three and a half hours"*.

*4.1.2 Active Authenticated Users*

Once registered and claiming, some people reported <u>difficulties using small-screens</u> to access UC Online stating it is "*quite frustrating... a bit of a fuddle"* and *"quite impossible on the phone because the screen is so small"* which can be a detriment to complying with commitments: *"if you can't read everything clearly, or if you miss something"*. Small-screens cause further difficulties when important correspondence, in the form of PDF letters, are only available online: *"I had to then get that downloaded onto my phone and so that was a bit of a faf"* and another participant noted *"if they put [a] letter on the journal thread, they should send you that in the post as well"*.

There can be <u>uncertainties about online functionality</u> such as knowing *"what did what and where"*. The Journal is a catch-all for non-existent functionality elsewhere, and with a lack of familiarity, it can be used by mistake: *"I put it in the journal because I didn't realise"*. Seven interview participants commented about the inability to upload documents easily: *"they actually have to send [a special link] via your Journal"* and *"if you want to send them another... you can't... you don't have the opportunity to do that"*. Data entry through the Journal becomes *"just one long stream of [unstructured] information"* which is *"really complex when you're trying to get hold of people"*.

Other <u>barriers to data provision</u> highlighted included limited topic options for changed circumstances reporting: *"there was no in-built change of circumstance [for my particular event]"* and *"you have to pick the best [matching] scenario"*. Some form inputs are reported as being overly restrictive: e.g.: *"I think it should be a bit more open,.. more free text"* and *"you can't log everything because it's not possible"*. Information such as job search reporting is *"laborious and monotonous"*. When sufficient information is not requested by forms or cannot be entered, this leads to additional follow-ups e.g.: *"they'll come back to you then asking for different documents... a bit annoying and can lead to delays"*.

### 4.2 Temporality

Many claimants mentioned issues caused by a mis-match between the service's timings and people's own lives. This led to six codes placed in a *Temporality* theme with two sub-themes: *Moment* reflecting timeliness when required, and once initiated, *Rhythm* of the subsequent interactions.

*4.2.1 Moment*

People's own life experiences are not predictable e.g.: *"I'd actually ended up homeless"*, *"your mental health... goes up and down"*. People expressed feelings about a <u>disjoint with their life experiences</u> where a rigidity does not match their own needs and circumstances: *"I thought I'd probably have a week... to report my changes and things"*. UC's interaction points are driven by notification alerts prompting claimants to log in at short notice or face sanction: "*by the time I got home, it was like six o'clock at night... it was something which needed, an action in quite urgently*", and "*we asked you to do something... so we are now going to sanction you*". People can contact the UC telephone helpline for some matters but they reported the long queue can interrupt other life activities.

Forced to use the Journal to send messages (responded to later), there is a broadly noted concern about <u>practical real-time interaction</u> at the time when needed e.g.: *"there's no interaction with them basically"*. Others expressed a desire for on-demand *"instant advice/help"* digital contact, with 16 participants suggesting using existing commercial video/screen-sharing services or webcams or phone cameras but with a recognition *"everyone is different some people would rather see someone face to face, talk on the phone, text/chat"*. There is a belief services should <u>proactively identify beneficial matters:</u> *"to be proactive"* that would help claimants in a timely manner in advance. This includes help *"with any additional benefits you may be entitled to and not know about"*, rather than after the event: *"a phone call, saying 'you didn't inform us'"*.

*4.2.2 Rhythm*

Claimants repeatedly commented on their frustrations about anything restricting their ability to progress a matter once started, with <u>slow responses</u> mentioned frequently e.g.: *"you're waiting four or five days and the situation is getting worse... that's creating all sorts of anxiety issues worry issues"*. Other impacts include payment delays, making it difficult to plan or *"budget ahead"*, and because *"people rely on uc [payments] being on time"* and *"that could be the difference between someeone eating or being evicted"*. Claimants are frustrated by the time to receive replies to their Journal queries e.g.: *"you type it in on your journal it's not done like real time"*, *"it can take a while for someone to message you back"*; and lack of response e.g.: "*sometimes they just ignore them*". Claimants want *"a more in depth like conversation... to get more accurate help"* to resolve matters quickly.

The detriments caused by these delays often fall on the claimants alone: *"they didn't get back to me... and then they tried to sanction me... because I didn't notify them soon enough"*. The <u>time taken to get decisions and status updated</u> is equally concerning to some e.g.: *"it took them 3 or 4 months to process that"*. Once decisions are made, people may not be informed e.g.: *"there was no pre-announcement"* and *"nobody's notified me"*. Two described their inability to have a particular status changed over many months: *"it would have gone on forever if I had not gone on insisting about it"*.

**4.3 Transparency**

This third theme, *Transparency*, is drawn from citizens' experiences about lack of insight of their rights, the systems and the processes being undertaken. Two sub-themes: *Actions* are principally related to matters of an ongoing interaction with the service (six codes), and *Ecosystem* expands this to the wider range of actors, mediating tools, rules and conventions that are required or may be used to support the provision and receipt of the service (eleven codes).

*4.3.1 Actions*

Claimants want greater <u>clarity in alert notifications</u> somewhat like ideas raised in scenario 2. The only method for claimants to view messages and reply is within the Journal. The email/SMS notifications inform claimants when there is a new Journal message but typically neither includes the topic, nor any indication of priority – one person stating they "*detest how UC sends you messages telling you to check your journal but cant show why*" which is *"really annoying and quite useless"* and other people disliking having to log into UC Online to read the content; there is not even a link to follow. This can delay action e.g.: *"think 'oh it is not urgent I will deal with it later'"*, or for others leads to worry and anxiety e.g.: *"I personally get really anxious when I get the check your Journal message"*. Although people expressed concerns about the risks of including confidential information, others would welcome it e.g.: *"the barest idea of what the journal message is about"* as some are generic notifications.

Participants also wish for better <u>transparency of decision-making processes</u>. Firstly, the status of actions e.g.: *"how long the [agent] actually spends on the portal... what he actually does"*. Secondly, how information is used e.g.: *"there's a lot of complexity"*, and particularly payment calculations e.g.: *"show the calculation"*. The participants welcome the Journal's permanent <u>record of interactions</u> e.g.: *"one bonus... it's documented"* and *"if it's on my Journal in writing they can't say 'I did not say that'"*, in contrast with UC's non-digital secondary channels e.g.: *"things getting lost, no proof, whereas the journal is proof"*. People expressed concerns that the telephone helpline has less accountability e.g.: *"I think your phone calls weren't even logged"*. People also want confirmation of information receipt and some need to extract data but there is no way to do so.

*4.3.2 Ecosystem*

Frustrations arise when claimants undertake <u>unnecessary data entry activities</u>. Information seems to be available which is not to the immediate benefit of the claimant such as having worked, but not reported it. In contrast claimants have to do extensive data gathering and re-keying, even data the DWP itself holds. Likewise, claimants' GP practices provide paper Fit Notes which claimants have to key into the UC website and take or send the paper to the JobCentre for validation, making people wonder *"why doesn't the doctor just send it straight off then?"*. Tenancy agreements from landlords were mentioned similarly.

There is a recognition that *"a lot of people don't understand their rights"* or *"understand the regulations"*. Consequently there is often a reliance on <u>third-parties for advice, support and guidance</u> about benefit rights and how to use the service e.g.: *"Universal Credit were coming back with not much help"*, *"it would be helpful having someone on my side..."*. Some seek more formal sources e.g.: *"I had to... go and see a benefits advisor"* with others seeking informal advice e.g.: *"I have to ask people around me"*, *"I'd get my mum to do it, or um some days I would do it"*, as well as *"I was in contact with people... by social media"*. Participants have difficulties sharing relevant information with others e.g.: *"there's no way for anyone else to log into the system"*, *"my daughter helps me, so she she'll know everything, she can access all my account"* and *"a welfare advisor... so I had to give her the username and password and a memorable word"*. There is a belief by some this informal sharing of access is happening widely.

### 4.4 Proficiency

*Proficiency*, spanning another sixteen codes includes citizens' experiences of unintentional events, divided into two sub-themes: *Slips and Errors* combining mishaps, mistakes, misunderstandings and other minor errors, and *Competence* in delivering and receiving what is defined by regulation.

*4.4.1 Slips and Errors*

Participants want greater allowances for minor failings e.g.: *"if people try to claim and get something wrong, they might lose their existing claim"* and *"there seems to be no flexibility in the system"*. These bring *"into question your trustworthiness and honesty"* and *"if there is a mistake, it's really troublesome"*. People also make some errors by being forgetful or just not knowing what to do. Even if errors are noticed, it can be difficult or impossible to correct them e.g.: *"I wanted to 'query a payment' but I accidentally pressed 'report a Fit Note'... - it was [then] on my to do list... I had to ask my doctor for a Fit Note"*.

There is also call for equitable accommodations for failings. Mistakes are not just made by claimants e.g.: *"a few [agents] who've missed my health problems ... so at that point sometimes I have to interject"*. With requirements on claimants to attend appointments (mostly telephone), there is little flexibility offered: *"if I rearrange, I have to have a really good reason for it"* but agents can miss appointments without consequence e.g.: *"I got an apology back... which is fine, stuff like that happens... it's not because anyone's at fault it's just one of those things"*. The imbalance of power is also revealed when claimants have been forced to put considerable effort into questioning decisions, e.g.: *"he commented that I would have some deductions which I said was wrong"*, which, after much claimant effort, are eventually accepted without further challenge e.g.: *"when they got notice of the appeal, they just backed down"* and *"the sanction was removed after two weeks"*. One noted there is not *"the same level of accountability that I have... in my working life, which I find very frustrating"*.

*4.4.2 Competence*

Claimants have concerns with attitudes towards them – participants want *"to be honest"* but *"know the system is abused... but crack down on the cheats, not the honest"*. This can manifest itself as feeling untrusted or lazy. Some agents are considered to be very helpful but the experience varies, e.g.: *"she disappeared and I got somebody who was much more direct and less personal"* and *"depending on who you speak to they're gonna be really lovely or they can just be really horrible"*. This leads claimants to wonder about agents' motives e.g.: *"I don't feel they have got the claimant's interest"* and *"make it difficult for you"*. This concern applies more generally e.g.: *"they're just doing it... [to] try and save public money"* and *"UC... appears to be set up to do all it can to stop you getting payments rather than making it easy"*. This can be made worse by any lack of consistency, especially when having to interact with multiple agents e.g.: *"if you get the same person twice... that's amazing... so you don't have to say everything again"*. Agent ability and agency can also be inconsistent e.g.: *"half of the times the people... don't seem to know themselves what they are doing"* and *"I got through to [the helpline] who initially was very very understanding, but not very helpful"*.

These contribute to mixed feelings about trust: claimants have concerns how data may be used for other purposes e.g.: *"that information, like a journal entry, might be going to"*, or how that might change e.g.: *"the DWP want to bring in this law where they can basically look at claimants' bank accounts"*. Increased access to personal data, such as in scenarios 3 and 4, led to negative reactions e.g.: *"an invasion of privacy"*, *"I wouldn't trust them... no it's GDPR"* and *"how long before they like before they start using it to investigate us?"*. However, some expressed a feeling that other organisations providing services or help, as illustrated in scenarios 1 and 3, need to be approved in some way to ensure quality, and the official public sector organisation ought to provide this anyway. One final comment sums this up: *"personally I think I'd be happier if it came from, the Department of Work and Pensions - and I can't quite believe I've said that"*.

## 5  DISCUSSION

Claimants' knowledge and experience has led us to provide our own insights into what about the digital channel is causing anxiety, feelings of frustration and additional burdens. Policy decisions define what processes and data are required, thus having an impact on people's experience of welfare benefit services, but government policies can be implemented in more than one way. For example in our study, participant claimants identified how communication delays arise through the deployment of an asynchronous written messaging system within the website (and not via any other method or channel, digital or otherwise). Similarly, assistance for some from advisors or other helpers is difficult because there is no delegated/shared access. And as a final example, claimants raised the frustrations and difficulties experienced having to request in advance every time they had to upload a file as evidence – one upload link for one file at a time per written request. These are digital design choices rather than being inherent policy matters defined by laws or regulations, or the result of general bureaucracy.

Our findings distinguish concerns with the digital implementation method, separate to delivery [18,22,40], to focus on what Alston describes as "the ways in which new technologies might transform the welfare state for the better" [4]. And participants mention the advantages identified by others [1,14,43,67]: convenience, knowledge/information access, discretion, and some reduced bias. However, we see how simplified views of people's lives have increased complexity for claimants (like [62]) with UC Online suffering from a mismatch with the realities of people's intricate (as Graeber [41]) and unpredictable (as Birhane [6]) lives. Like noted by Morris et al. [62] and Considine et al. [21], people's lives are messy; we identified how UC Online is a barrier, affects people's wellbeing and does not provide people with sufficient agency to interact with the DWP, despite requirements to promptly report change of circumstances, and respond to Journal messages. This is what Reeves [73] described as a system "beyond their control". We see how a focus on internal service efficiencies overlooks overall societal efficiency; the transfer of burdens to claimants and their own support networks also needs to be accounted for by the State. In particular, we identified issues of temporality, not addressed elsewhere, where the mismatch between people's life experiences clashes with the system, and the pace of subsequent interactions does not fit with people's needs or expectations. The ongoing and remote nature of digital interactions distinguishes UC from previous work on non-digital systems [92]. We also expose a lack of trust (as Alessiato [1], and Scullion and Curchin [78]) which our participants believe could be partly addressed by greater transparency about current status and how decisions are made, combined with being treated better (like reciprocity in Coles-Kemp et al. [18]) by the agents, who are interacting facelessly [62] through the digital system. Our participants repeatedly point to examples of better designs they have seen in other digital systems. As noted previously, others [15,88] identified how UC Online has increased claimants' support needs. Our study has exposed how the digital channel itself suppresses the ability for information sharing with, and getting help from other people and groups in claimants' own socio-technical ecosystems.

We have developed eight recommendations, based on the identified insights from the study's participants in a hierarchy of three categories. They reflect how service provision should not be a process, but rather accommodate the realities of people's lived experiences. We conclude with points of note regarding limitations, potential negative impacts, and future work.

### 5.1 Implications for design

We provide recommendations for inclusive and just design practices [28] for digital technologies providing welfare benefits services for people in a precarious low-income situation. These are regardless of whether the systems belong to a public sector organisation, a third party organisation or business, the individuals themselves or their communities [22].

*5.1.1 Supporting claimants' own socio-technical ecosystems*

Massimi et al. [58] and Semaan's [79] explain how life disruptions help expose failings in infrastructure. This is particularly the case for welfare benefits, where Harris [47] has pointed out the need for a "degree of complexity in relation to social security provision" to safeguard welfare rights, and therefore not all claimants can ever be fully self-reliant. Thus there is a need to help people achieve their desired outcomes (as [29,92]), and systems should acknowledge and facilitate the leverage of formal and informal third-party help and resources to increase people's "everyday resilience" [79]. These provide various advice about benefit rights, ongoing support to access and use systems and guidance information. Our participants did not need "intermediated technology use" [77], but instead use existing support assets when required (as suggested by Pei and Nardi [69]) since their needs vary with their life patterns, as Massimi et al. also observed [58]. There are costs to those support networks, as Pei and Crooks noted [68], and we have found that digital implementation choices can make these network costs greater, while the State attempts to achieve internal service delivery efficiencies. Digital service provision can isolate people further from community assets, and we found how design choices can make it more difficult for claimants to get help from support networks. These support networks are not just Dombrowski et al.'s intermediaries [29], but also friends, family and individuals/groups in the wider community who act together with claimants. We also identified how a lack of transparency further adds to costs transferred to these support networks. Our two related design implications are:

Recognise how wider networks contribute to welfare benefit service delivery: For welfare benefit systems, the "service user" is rarely just one person and there should be support for other actors in people's wider ecosystems. Facilitate and support these networks and existing assets by enabling, encouraging and promoting information sharing, delegated access, real-time-integrations, and other ways to involve the help of others.

Generate structured case data to support visibility and re-use by claimants and other parties: Provide, signpost and explain functionalities to span the fullest range of claimant activities supported, so as to reduce the need for and use of messaging/chat interactions (like the UC Journal) which result in unstructured data. Ensure the structured data supports re-use to increase wider network efficiencies: provide for recording and keeping full records of every interaction by every channel (including files, form submissions, all communications), and provide methods to ensure information can be found and referred to, and easily exported or shared with others as required.

*5.1.2 Acknowledging welfare benefit service users as people*

We see how digital technologies can exacerbate power differences between welfare benefit claimants and the State: *"you have to react and sort of cow down to what they want".* There is an imbalance of power (as [50]) and mistrust (as [14]), with concerns that digital systems are not working in claimants' interests, and unaccountable (as [65]) remote agents are able to use discretion against them (as [70]). Public service digitisation is known to be at risk of "weakening professional and relational values" [12] through reduced personal interaction, and our study found concerns with attitudes towards claimants who desire respect, to be treated as honest if imperfect (as [50] and like Coles-Kemp et al.'s "reciprocity and care" [18]), rather than being assumed to be dishonest and lazy. There is sense of a lack of transparency of decision-making processes including payments (also noted by Carey and Bell [14]), and what evidence was used and how. Our findings contribute to observations on other welfare benefit jurisdictions' fair use of data as in [1,50,55] and how digital technologies can lead to the "co-destruction" of value as in [37]. However, despite a high level of concern by some participants, the same people thought the relevant public sector organisation should be the primary source for information, help, tools and decisions. Our three related design implications are:

Prioritise claimants' interests over system efficiencies: All processes, methods and decision-making should prioritise claimants' needs to achieve best outcomes for individuals rather than system efficiencies. Organisational knowledge and resources should be utilised to this respect including intervening in advance to identify matters that affect claims or what claimants may have forgotten about.

Ensure system and State accountability to claimants: Equalise accountability between claimants and the State. Promote a sense of fairness by enforcing an expectation that service level standards for actions and response times should be similar to those expected of claimants, with related penalties not disproportionately, or only, affecting claimants. Provide tools/methods for claimants to easily check, query and challenge actions and decisions.

Provide clear and configurable communications about process and decision statuses: Use methods proactively, such as internal notifications and external alerts, to help people understand what they need to do/when, and what is in progress by others. Provide confirmations when actions have been completed, information received, decisions made and statuses changed. To accommodate people's individual needs and preferences, provide choices about what, when and how these are received, who they are sent to and copied to. Offer content options such as whether to have a topic, priority and hyperlink, and ensure it is clear where and how to complete any required action.

*5.1.3 Reducing claimants' burdens of system interactions*

The need for assistance from the State coincides with when people's cognitive capacity is reduced, limiting access to welfare benefits and thus increasing inequality [16] (as Sin et al. [84] found with older adults). With a digital channel being used to both access and deliver the service, barriers then continue to be present as people maintain their welfare benefit awards and receive payments. Claimants' difficulties are worsened (as [73]) by an unfamiliarity with benefit rights, regulatory requirements and service provision, leading to greater uncertainties about online functionality: what to use, when, how and where. Limitations with interfaces, including lack of functionality, reflect service journeys which do not match life experiences. Similar to Holten Møller et al. [49], claimants do not want to do what they perceive as unnecessary work such as data entry activities (being an active data provider does not provide real agency, like Holten Møller et al. [50]). However, we found claimants instead want insight and control over data about them (which [49] noted is particularly challenging when there is increased reliance on data). Different time, different place, or asynchronous distributed interactions [36], lead to a disjoint with claimants' life experiences (as Cheetham et al. [15]). Our participants repeatedly identified this as contrary to their experiences with synchronous interactions in other systems online, and also across other channels (like multi-touchpoint experience design [76]). Our final three further related design implications are:

Shift the burden of gathering evidence from claimants towards the State: Transfer effort from claimants to the State, to improve timeliness and reliability. Prioritise the implementation of digital processes to gather or import, check and use necessary data from internal and external providers. Change claimants' role to verifying the data, rather than providing it, and ensure claimants have visibility and control over derived status attributes.

Provide greater flexibility and accommodations for claimants in the accuracy, precision, timeliness and permanence of the remaining information they provide: Design for people's lives which can be complex and where changing events and circumstances, often beyond their control, drive their need and eligibility for welfare benefits services. Allow adequate, rather than complete and precise, data that suffice for the State's needs. Increase flexibility of use by avoiding strict deadlines; limiting the use of actions that block progression; permitting correcting, updating and reversing information; and withdraw penalties for simple slips and lapses.

Provide full welfare benefits service across wider interoperable channels: Ensure all service provision and modes of help (advice, support and guidance) are available through multiple interaction channels (e.g.: telephone, web, mobile app) which are accessible to varying resources and capabilities (e.g.: communication skills, equipment, language, physical and mental abilities). Permit the use and intermixing of channels without restriction. Consider providing on-demand synchronous interactions through digital as well as other channels.

**5.2 Limitations, wider impacts and future work**

Our participants described their own individual experiences, some common to multiple participants, but there is no intention they are representative of most people's experiences. Our sample was constrained by an inability to engage with potential partner organisations affected by the pandemic (as others found [35]), by being less able to contact those with greater digital exclusion issues, and by claimants' wariness which seemed to be due to a lack of institutional trust in the DWP, the public sector organisation responsible for UC. The timing also meant employment related commitments were not being enforced, and so aspects like job search and job placement have not been exposed by our study (see instead, for example, work in another jurisdiction [50,51]). Despite this some participants had not ceased *"I've just been doing that anyway, because I'd prefer not to be on Universal Credit"*. Although based on a study of one system, our design recommendations are presented more generally, relevant to service provision to this common group of precarious working-age adults. Our design implications are intended for researchers and designers. The authors themselves are currently developing prototype tools using participatory design practices to apply our design implication recommendations. There is a recognition that insights could also be used against, rather than in the interests of claimants, and further work looking at similar systems needs to be aware of the wider human rights issues and risks.

## 6 CONCLUSION

This paper adds to the understanding of claimants' experiences of an online welfare benefit system to take up their rights, by a study of the UK working-age benefit Universal Credit using remote interviews (n=11) and online surveys (n=66). These capture aspects of the vital mismatch between a process driven service on the one hand and the realities of people's lived experiences on the other, where many individuals are reluctant or involuntary users. We find that digital channels can contribute to additional frustrations, anxieties and burdens for this group of people who want to be acknowledged as people. There is a need for digital channels to actively support and bring together the combined capabilities of individual citizens, the wider community and the State, and to do so in a timely manner for each individual's needs. Our contributions are these insights from the study's participants and design implications to inform further research and provide guidance for ongoing and future inclusive and just design and development of digital technologies providing welfare benefits services for people in a precarious low-income situation. These aim to support the development of less harmful, and more beneficial systems which might impact not only claimants like our participants, who are able to use such online systems, but also those less able or unable to do so.

## ACKNOWLEDGMENTS

The authors wish to thank the workshop participants for their contributions to the research activity. This research was funded by the EPSRC Centre for Doctoral Training in Digital Civics (EP/L016176/1). Data and materials created during this research are available at https://doi.org/10.25405/data.ncl.c.6746076 in Newcastle University's data repository.